\documentclass[11pt,twoside]{article}

\usepackage{asp2004}
\usepackage{epsf}
\usepackage{psfig}
\usepackage{lscape}

\begin{document}
\title{ASTRA Spectrophotometer: Reduction and Flux Calibrations}
\author{Barry Smalley$^{1}$, Austin F. Gulliver$^{2}$, Saul J. Adelman$^{3}$}   
\affil{
{$^1$Astrophysics Group, Keele University, Staffordshire ST5 5BG, UK}\\
{$^2$Department of Physics \& Astronomy, Brandon University, Brandon, MB R7A 6A9, Canada}\\
{$^3$Department of Physics, The Citadel, 171 Moultrie Street, Charleston, SC 29409, USA}\\
}

\begin{abstract}

The ASTRA Cassegrain Spectrophotometer and its automated 0.5-m telescope at
Fairborn Observatory in Arizona will produce a large quantity of high-precision
stellar flux distributions. A separate paper \citep{ADE+07} presented a review
of the design criteria for the system and an overview of its operation. This
paper discusses the techniques used in the data reduction to final flux
calibrations.

Extraction of 1-d spectra from the 2-d images will be performed by a highly
automated version of CCDSPEC \citep{GH02}. The characteristics of the CCD are
automatically applied to the images, including the location of dead rows and
hot pixels. In order to achieve the goal of better than 1\% precision, large
numbers of bias and flat field frames will be used in the reduction process.
There will be a continual programme to monitor the image quality. Finally,
optimally extracted spectra will be obtained, including the removal of
scattered light and cosmic rays.

The Earth's atmosphere has a considerable effect on the stellar flux as
measured from the surface. The principal sources of extinction, Rayleigh and
aerosol scattering, ozone and telluric line absorption, are discussed, along
with methods used to determine their effects on the observed spectra.
Correction for telluric lines is the most problematical, due to their
non-linear variation with airmass. By using a large network of constant stars
to monitor atmospheric extinction it is possible to determine the extinction
coefficients to generally better than 1\% and to assess their temporal
variability.

The spectrophotometric observations are placed on an absolute flux scale by
reference to stars with known values of true flux at top of Earth's atmosphere.
These standard stars have been calibrated against terrestrial sources of known
properties. Unfortunately, very few stars have been calibrated directly. The
ASTRA fluxes will be calibrated against the best available Vega flux
distribution. The constant stars used in the extinction determinations will
provide the internal calibration network of secondary flux standards.

The available absolute calibrations are accurate to typically 1--2\%.
Ultimately this uncertainty will limit the accuracy of the final fluxes of
other stars. However, the internal precision will be significantly higher, and
should more-accurate absolute calibrations become available the fluxes can be
re-calibrated to higher accuracy.

\end{abstract}

\section{Introduction}

The ASTRA Cassegrain Spectrophotometer and its automated 0.5-m telescope at
Fairborn Observatory in Arizona will produce a large quantity of high-precision
stellar flux distributions. A separate paper, \citep{ADE+07}, presented a
review the design criteria for the system and an overview of its operation. In
this paper we discuss the techniques used in the data reduction to final flux
calibrations.

ASTRA will produce a vast quantity of high quality data. The reduction
procedures will be automated as far as possible, using software sentinels to
watch for unusual events.

\section{Reduction of 2-d Images}

The spectrophotometric observations taken by ASTRA comprise CCD frames
containing two spectral orders. The orders are sufficiently separated that
there is no order overlap and each order contains both the target spectrum as
well as sky background. Spectral extraction is performed using a highly
automated version of the CCDSPEC reduction package developed by \cite{GH02}.

Spectral Extraction follows the procedures common to normal spectra, starting
with correction for CCD properties, including dead rows, hot pixels. Next, bias
corrections and flat fielding is performed using master calibration frames
obtained from the addition of many bias and flat field frames. Order tracing is
performed and PSF fitting using optimal extraction, followed by scattered light
removal and the removal of cosmic rays. Details can be found in \cite{GH02}.

As part of the commissioning process we will characterise the extent of CCD
fringing, determine shutter timing corrections, and assess the nature of any
flexure of spectrograph.

Unlike normal spectroscopy, spectrophotometry can be regarded as precise
relative photometry at many wavelengths. Hence, the spectral extraction
procedure must recover exactly the same relative proportion of observed flux at
every wavelength from all observed spectra. Since the spectral width depends on
atmospheric seeing and the amount of telescope trailing, the extraction
aperture must be varied to ensure the same fractional spectral energy is
covered. The \cite{HOR86} optimal extraction algorithm maintains
spectrophotometric calibration validity.

It is also important to preserve the observed count level and not re-bin onto a
linear wavelength scale, since this will affect the atmospheric extinction
determination. The output from this stage of the reduction procedure is
instrumental wavelengths and counts/s per wavelength bin, plus an estimate of
the uncertainty in the counts.

\begin{figure}[!ht]
\plotone{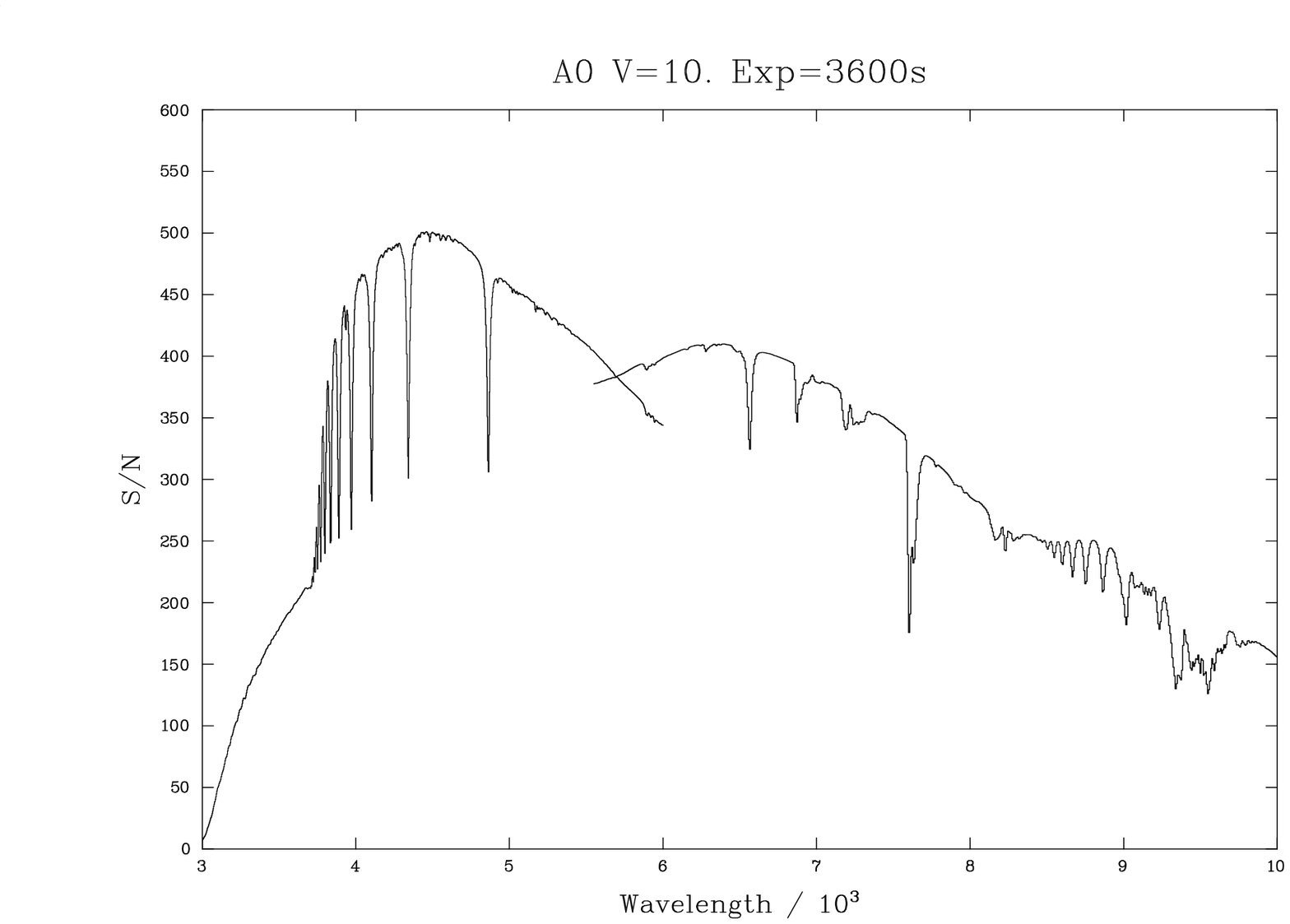}
\caption{Simulated Extracted Spectrum. This is given in estimated S/N for a
3600-second exposure of a 10th magnitude A0V Star.}
\label{smalley-fig1}
\end{figure}

Figure~\ref{smalley-fig1} shows an example of a simulated extracted spectra for
a 10th-magnitude A0V star. It shows that the 1-hour integration has produced a
signal-to-noise (S/N) in excess of 200:1 in most of the range from
$\lambda\lambda$3700--9300.

\section{Earth's Atmosphere}

The Earth's atmosphere has a considerable effect on the stellar flux as
measured from the surface. The principal sources of extinction, Rayleigh and
aerosol scattering, ozone and telluric line absorption, will be discussed,
along with methods used to determine their effects on the observed spectra.

\subsubsection{Rayleigh Scattering}

Rayleigh scattering due to molecules in the atmosphere is proportional to
$\lambda^{-4}$, with detailed formulae given by \cite{ALL73}, \cite{HL75} and
\cite{FS80}. Vertical extinction is proportional to local atmospheric pressure.

\subsubsection{Aerosol Scattering}

Aerosol scattering is due to dust particles, salt particles, water droplets,
man-made pollutants in the earth's atmosphere \citep{HL75, BUR+95, FOR+96}. We
can model aerosol absorption using Angstr\"{o}m's simple empirical
approximation formula $A_0 \lambda^{-\alpha}$. Aerosol extinction is quite
variable, with $A_0$ showing diurnal and seasonal variations to a factor of two
or more, and $0.5 \la \alpha \la 1.5$.

\subsubsection{Ozone bands}

The Huggins band provides the ultraviolet cut-off and the Chappuis band is
present in the green part of the optical spectrum, with two strong diffuse
peaks around 5730\AA\ and 6020\AA. High-resolution absorption coefficients are
available \citep{BUR+99, VOI+01}. Ozone is concentrated at altitudes between 10
and 35 km, and can vary significantly in a few hours \citep{HL75}. Typically,
Chappuis can contribute around 0.01~{mag.} of absorption per airmass.

\subsubsection{Telluric lines}

Several bands of discrete absorption lines, especially in the red and near-IR:
oxygen (O$_2$) around 7590\AA\ and 6870\AA, and water vapour (H$_2$O) bands
near 7100\AA, 8090\AA, 8920\AA, and 9277\AA. Water vapour is highly variable
with irregular night-to-night and seasonal variations. The H$_2$O bands do not
follow the normal exponential absorption (Bouguer) law, but show a
curve-of-growth effect \cite{HAY70}.

\begin{figure}[!ht]
\plotone{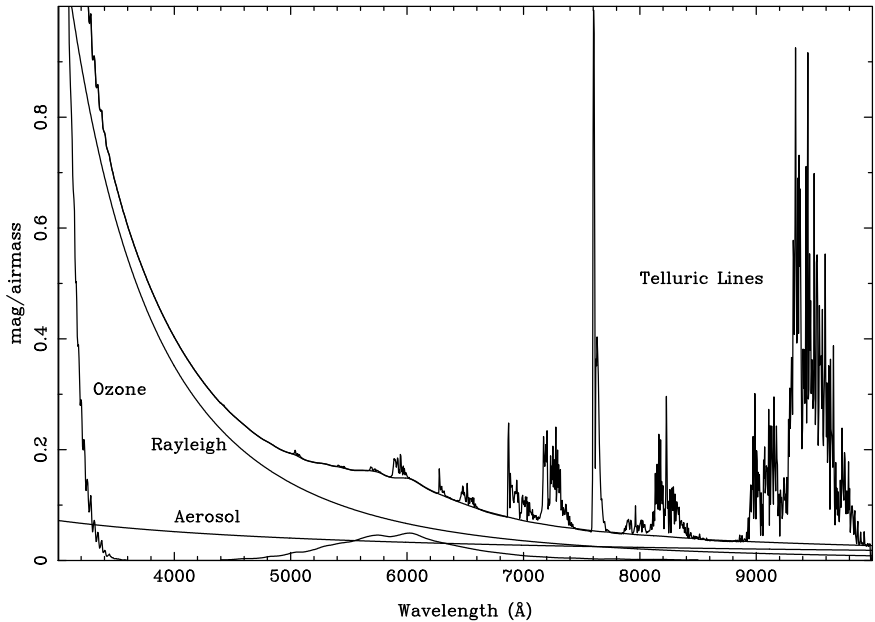}
\caption{Simulation of the typical extinction by Earth's Atmosphere, showing
the relative relative contributions of Rayleigh and aerosol scattering,
ozone and telluric line absorption.}
\label{smalley-fig2}
\end{figure}

\subsection{Extinction Law}

Extinction by the Earth's atmosphere is given by Bouguer's Law,
\[ m_\lambda = m_{\lambda0} + k_\lambda X, \]
where $m_\lambda$ is observed stellar magnitude at wavelength ($\lambda$);
$m_{\lambda 0}$ is stellar magnitude above Earth's atmosphere; $k_\lambda$ is
extinction coefficient (mag./airmass); and $X$ is the airmass relative to that
at the zenith.

The total extinction coefficient at each wavelength is the sum of the various
individual contributions: Rayleigh, Aerosol, Ozone, Oxygen and Water (see
Fig.~\ref{smalley-fig2}).

\subsection{Airmass}

In the classical text, \cite{HAR62} gives a polynomial formula for calculating
airmass ($X$) from zenith distance ($z$). In his review, \cite{YOU74} discussed
airmass calculations and advocated the use of a simpler formula,
\[X = \sec z [1 - 0.0012 (\sec^2 z -1) ],\]
which is valid up to $\sec z = 4$ and good to better than 1\%. \cite{YOU94}
also warns that airmasses greater than 4 should be avoided because of large
random and systematic errors.

Airmass can change significantly during a long exposure. In such cases we can
use effective airmass \citep{STE89},
\[ X_{\rm eff} =
  \frac{1}{6} \left[X_{\rm START} + 4 X_{\rm MIDDLE} + X_{\rm END} \right]. \]
This is valid provided that extinction is not varying significantly as during
the exposure. An alternative is to take multiple shorter exposures.

\subsection{Seeing}

In slit-less spectroscopy resolution depends on seeing. We can characterise the
spectral resolution as follows,
\[ \frac{\Delta\lambda}{\lambda} \propto
\sqrt{\theta^2_{\rm seeing} + \theta^2_{\rm inst}}, \]
where $\theta_{\rm seeing}$ is the size of the stellar seeing disk and
$\theta_{inst}$ is the fixed instrumental resolution for an idealised
point-size stellar image.

Seeing is naturally highly variable, with not only temporal variations
($\theta(t)$), but also variations with airmass ($\theta \sim\ X^{0.6}$) and
wavelength ($\theta \sim \lambda^{-0.2}$) \citep{WOO82}. The variation with
wavelength can be regarded as fixed, but this resolution is different from the
instrumental value as determined by arc spectra.

\subsection{Spectral Resolution Issues}

\begin{figure}[!ht]
\plotone{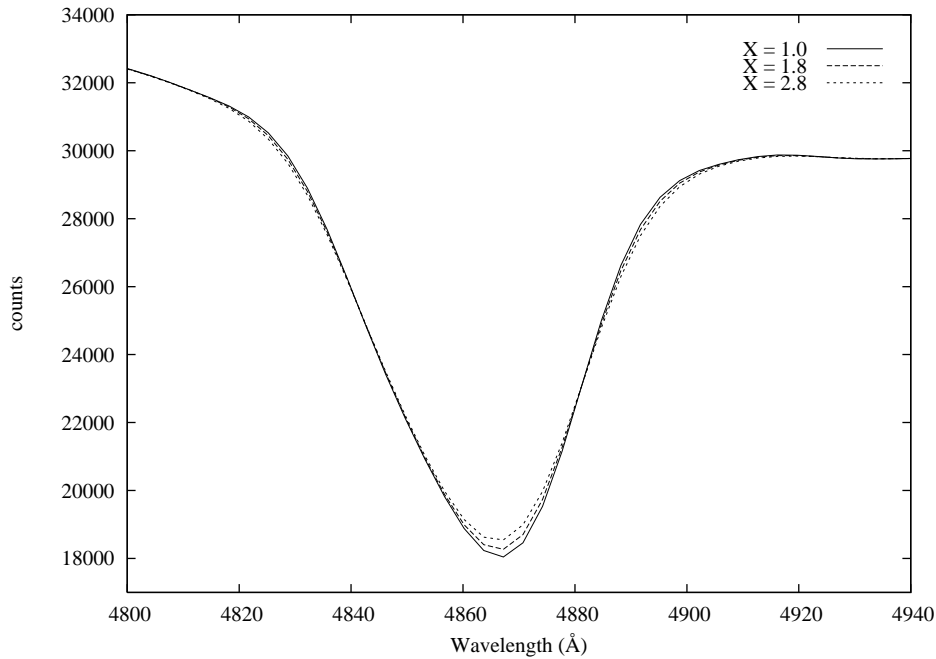}
\caption{The effect of resolution varying with airmass on the shape of H$\beta$.
The core is noticeably affected, and note the more subtle effect in the wings.}
\label{smalley-fig3}
\end{figure}

Since the resolution of the spectra is variable, we must allow for this during
the extinction determination. This is most important where spectrum is varying
most rapidly, e.g. Balmer lines. Table~\ref{smalley-table} gives the results
that are obtained when simple Bouguer law is used to fit to various points
within the H$_\beta$ Balmer line during a simulated night of relatively poor
seeing. The fit for the Balmer core is significantly different from the true
value, as can be seen from Fig.~\ref{smalley-fig3} which shows that the depth
of the core varies noticeably with airmass.

\begin{table}[!ht]
\caption{Extinction determined from Bouguer Law fits at various wavelengths
within the H$_\beta$ Balmer lines compared to the actual value.}
\label{smalley-table}
\smallskip
\begin{center}
{\small
\begin{tabular}{ccc}
\tableline
\noalign{\smallskip}
$\lambda$ &  $k$(fit) & $k$(true) \\ \hline
\noalign{\smallskip}
\tableline
\noalign{\smallskip}
4867 & 0.157 & 0.184 \\
4894 & 0.178 & 0.180 \\
4940 & 0.174 & 0.176 \\
\noalign{\smallskip}
\tableline
\end{tabular}
}
\end{center}
\end{table}
The effects of varying resolution become important on nights of poor seeing and
when seeing is variable.

\subsection{Scintillation}

Scintillation noise is due to motion in the Earth's atmosphere during an
exposure and sets a limit on photometric accuracy. This can be estimated using
the \cite{DRA98} formula,
\[ \sigma =
   0.09 D^{-2/3} X^{1.75} \exp \left(\frac{h}{H} \right) (2T)^{-1/2}, \]
where $\sigma$ is the scintillation noise in magnitudes, $D$ is the diameter of
the telescope (cm), $h$ altitude of the telescope (m), $H$ scale height of the
atmosphere ($\sim$8000m) and $T$ is the exposure time in seconds.

The normal minimum exposure time for ASTRA is 10 seconds in order to minimise
scintillation noise. A 10-second exposure gives $\sigma$ = 0.001 mag. at the
zenith and 0.007 mag. for X = 3. Longer exposures give lower levels of
scintillation noise.

\section{Extinction Determination}

Extinction determination will be performed without the use of any assumed
stellar true fluxes. The extinction stars are tested to be constant, with any
variability $\ll$ 0.01 mag. Any extinction stars found to be significantly
variable will be removed from the list of standard stars and extinction
re-determined. Extinction determination is performed for a single night of
observations and its reliability assessed. We will investigate the use of
multi-night reductions.

The brightest stars have to be observed through a neutral density filter. We
will determine and monitor the throughput. The filter transmission function
enters the extinction determination as an extra term in the fitting procedure
\cite{SM92}.

Atmospheric extinction is variable. This usually appears as a gradual decrease
during the night, due to a slow fallout of aerosols \citep{YOU74}. For a few
hours, the extinction change can be represented by a linear function with time.
Observationally, we must observe enough extinction stars to maintain a nearly
continuous check on the extinction coefficients. Short exposures are required
to avoid significant changes in airmass and extinction.

\cite{YI67} weighted each observation by $1/\sec z$ so that residuals have
units of mag./airmass. Thus residuals against time allows deviations from mean
extinction to be spotted.

Initially we assume a perfect night and use a time-independent multi-star
Bouguer Law fitting procedure as described in \cite{SM92} using least-squares
techniques \citep[Chap.~15]{PRE+92}. The residuals from the fits are assessed
using a variety of techniques to look for variations and correlations with time
and/or airmass. Checks are also made to ensure that none of the extinction
stars is variable. As well as examining the $\chi^2$, we will use R-statistics
as defined by \cite{BS93},
\[ R =
  \frac{1}{\sum \sigma^2 (n-1)} \frac{\sum y_{i} y_{i+1} w_i}{\sum w_i/n}, \]
where $w_i = \frac{1}{\sigma_{i} \sigma_{i+1}}$. These measure the degree of
correlation of the residuals for adjacent values, which must be strictly
monotonic.

\begin{figure}[!ht]
\plotone{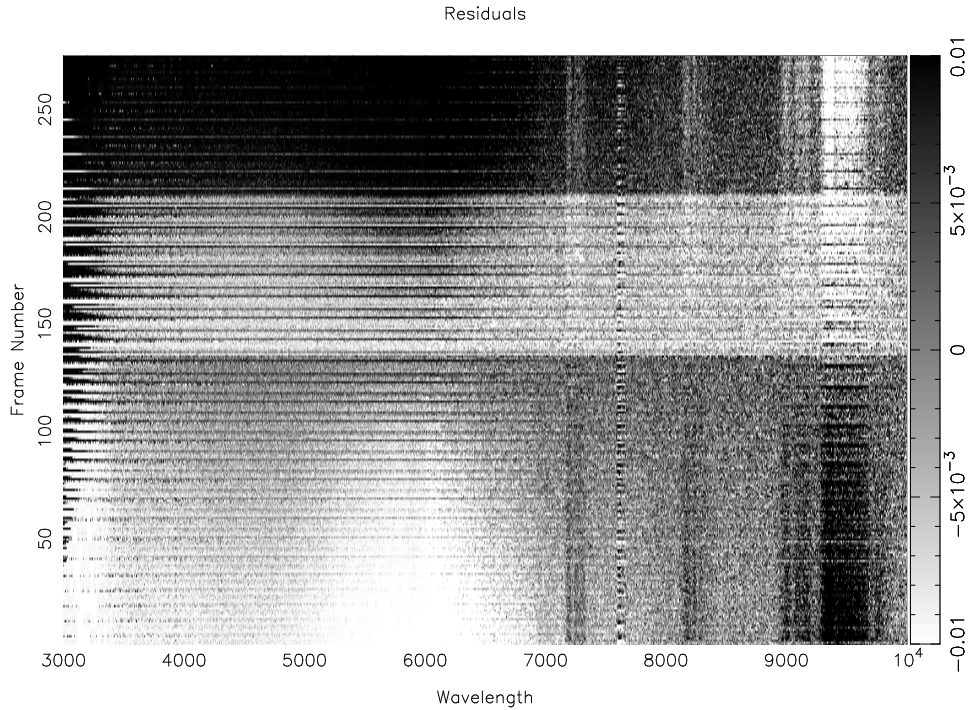}
\caption{Simulation of a night with variable extinction. The residuals from the
extinction determination reveal obvious variations in ozone and
water vapour, as well as a spell of grey extinction due to a
simulated cloud band between frames 140 -- 210. The image is ordered by frame number
as a proxy for time.}
\label{smalley-fig4}
\end{figure}

In addition, a 2-d grey-scale image plot of residuals against wavelength and
time is a very useful diagnostic. For example, Fig.~\ref{smalley-fig4} shows a
simulation of a night with highly variable extinction, including changes in
ozone and water, and a spell of grey absorption due to a passing cloud band.

If variable, we can use $k(t)$ as a low-order polynomial in time, adding extra
terms  and using an F-test for the significance of improvement
\cite[Chap.~10]{BEV69}. If necessary, we can split the night into smaller
blocks, in order to avoid parts with poor conditions.

\section{Fitting Telluric Regions}

Non-linear extinction violates Bouguer's Law. Several methods have been used in
the literature, but we will use synthetic telluric spectra. Non-linear
least-squares fitting is used to determine oxygen and water columns and their
temporal variability. Theoretical transmission functions can be calculated
using the HITRAN molecular line lists \citep{ROT+05} and a simplified 6-layer
model of the Earth's atmosphere \citep{NIC88}. These are calculated with
varying water and oxygen columns and tabulated at high resolution (0.5\AA\
steps) for rapid interpolation.

At the resolution of the ASTRA instrument the individual telluric lines are not
resolved. Thus, flux through each resolution element will depend on the nature
of the intrinsic telluric lines and any stellar features within that region.
Strong absorption (or emission) features may be hidden by the lines or
`visible' between them. We cannot {\it a priori\/} know which case is occurring
in each band \citep[e.g.][]{STE94}. We also have to beware of stellar lines
(dis)appearing in and out of telluric lines due to radial velocity variations.
The Earth's orbital velocity (30 km\/ s$^{-1}$) gives a 0.5\AA\ shift at
5000\AA. We will assess the level of this uncertainty using synthetic stellar
spectra.

\section{Flux Calibration}

Spectrophotometric observations must be placed on an absolute flux scale by
reference to stars with known values of true flux at top of Earth's atmosphere.
Standard stars have been calibrated against terrestrial sources of known
properties. Unfortunately, very few stars have been calibrated directly. The
primary standard is $\alpha$ Lyr (Vega). The standard references for absolute
fluxes are \cite{HL75}, \cite{TUG+77} and \cite{HAY85}. The accuracy of these
and other calibrations was assessed by \cite{MEG95}.

Most calibrations are at relatively low resolution (typically 50--100\AA).
Corrections for bandpass effects need to be considered, which may require the
use of high-resolution spectra. \cite{COL+96} presented a calibration for Vega,
for use with HST data, which used Kurucz ATLAS9 fluxes to generate a
higher-resolution. A more-recent absolute calibration based STIS observations
was determined by \cite{BG04}. A discussion of HST standards is given by
\cite{BOH07}.

The ASTRA spectrophotometry will be placed on the Vega absolute flux scale.
This will require the use of a neutral density filter, whose throughput will be
measured and periodically checked. Whenever, we use the neutral density filter
we will observe a suitable star with and without the filter to determine the
throughput.

Our quality control procedures will assess stellar (micro) variability and the
long term variability of extinction stars to ensure they are constant. In
addition, we will monitor the instrumental and telescope throughput for
variations. These tests will ensure flux consistency within secondary standards
and of the fluxes of all target stars. Re-calibration of whole data set be will
performed as necessary. For standards we will use only the best observations at
the highest resolution. For other observations, we will ensure spectral
resolution effects are allowed for.

Flux calibrated spectra for all observations will be produced. The individual
observations of non-standard stars will be co-added, if non-varying, in order
to increase the signal to noise. Variable stars will not be co-added, except if
the variations are insignificant between observations.

The reduction procedures will store intermediate results, such as
extracted spectra in counts, instrumental counts/second above Earth's
Atmosphere, details of the nightly extinction models, and individual flux
calibrated spectra.

The final fluxes will be made available as FITS files including headers and
processing history. For each spectral point, we will give wavelength, bin size,
instrumental (counts/s), flux relative to Vega, absolute fluxes
(photons/s/nm), internal errors, external errors, and quality flags.

\section{Conclusions}

The reduction of ASTRA spectrophotometry requires careful extraction and
calibration in order to achieve stellar flux measurements with internal
(star-to-star) precision better than 1\% for stars brighter than $\sim$10th
magnitude. In the regions heavily affected by telluric lines we will not
necessarily reach the 1\% precision level. For all observations, including
faint spectrophotometric standards, we will use full error propagation to
include both internal and external errors. 

The available absolute calibrations are accurate to typically 1--2\%.
Ultimately this uncertainty will limit the accuracy of the final fluxes of
other stars. However, the internal precision will be significantly higher, and
should more-accurate absolute calibrations become available the fluxes can be
re-calibrated to higher accuracy.

\acknowledgements 
This work is supported by NSF Grant AST-0115612 to The Citadel, Saul J.
Adelman, Principal Investigator. This paper is ASTRA paper number 6.

\end{document}